\documentclass[journal]{IEEEtran}
\usepackage{cite}
\usepackage{amsmath,amssymb,amsfonts}
\usepackage{algorithmic}
\usepackage{textcomp}
\usepackage{algorithm}
\usepackage{algorithmic}
\usepackage{mathrsfs}
\usepackage{booktabs}
\usepackage{graphicx}
\usepackage{subcaption}
\usepackage{longtable}
\usepackage{hyperref}
\usepackage{cleveref}
\usepackage{verbatim}
\usepackage{xcolor, soul}
\sethlcolor{yellow}

\begin{document}

\input epsf



\newtheorem{lemma}{Lemma}
\newtheorem{corol}{Corollary}
\newtheorem{theorem}{Theorem}
\newtheorem{proposition}{Proposition}
\newtheorem{definition}{Definition}

\title{Integrated Sensing and Communication based Outdoor Multi-Target Detection, Tracking and Localization in Practical 5G Networks}

\author{Ruiqi Liu, Mengnan Jian, 
Dawei Chen, Xu Lin, Yichao Cheng, Wei Cheng and Shijun Chen 
\thanks{The authors are with the Wireless Research Institute, ZTE Corporation, Beijing 100029, China and the State Key Laboratory of Mobile Network and Mobile Multimedia Technology, Shenzhen 518055, China (emails: \{richie.leo, jian.mengnan, chen.dawei2, lin.xu1, cheng.yichao, cheng.wei79, chen.shijun\}@zte.com.cn).}
\thanks{Corresponding author: Dawei Chen.}}

\maketitle

\begin{abstract}
The 6th generation (6G) wireless networks will likely to support a variety of capabilities beyond communication, such as sensing and localization, through the use of communication networks empowered by advanced technologies. Integrated sensing and communication (ISAC) has been recognized as a critical technology as well as a usage scenario for 6G, as widely agreed by leading global standardization bodies. ISAC utilizes communication infrastructure and devices to provide the capability of sensing the environment with high resolution, as well as tracking and localizing moving objects nearby. Meeting both the requirements for communication and sensing simultaneously, ISAC based approaches celebrate the advantages of higher spectral and energy efficiency compared to two separate systems to serve two purposes, and potentially lower costs and easy deployment. A key step towards the standardization and commercialization of ISAC is to carry out comprehensive field trials in practical networks, such as the 5th generation (5G) network, to demonstrate its true capacities in practical scenarios. In this paper, an ISAC based outdoor multi-target detection, tracking and localization approach is proposed and validated in 5G networks. The proposed system comprises of 5G base stations (BSs) which serve nearby mobile users normally, while accomplishing the task of detecting, tracking and localizing drones, vehicles and pedestrians simultaneously. Comprehensive trial results demonstrate the relatively high accuracy of the proposed method in practical outdoor environment when tracking and localizing single targets and multiple targets.
\end{abstract}


\section{Introduction}\label{S1}

\begin{figure*}[t] 
            \centering
            \includegraphics[width=0.8\linewidth]{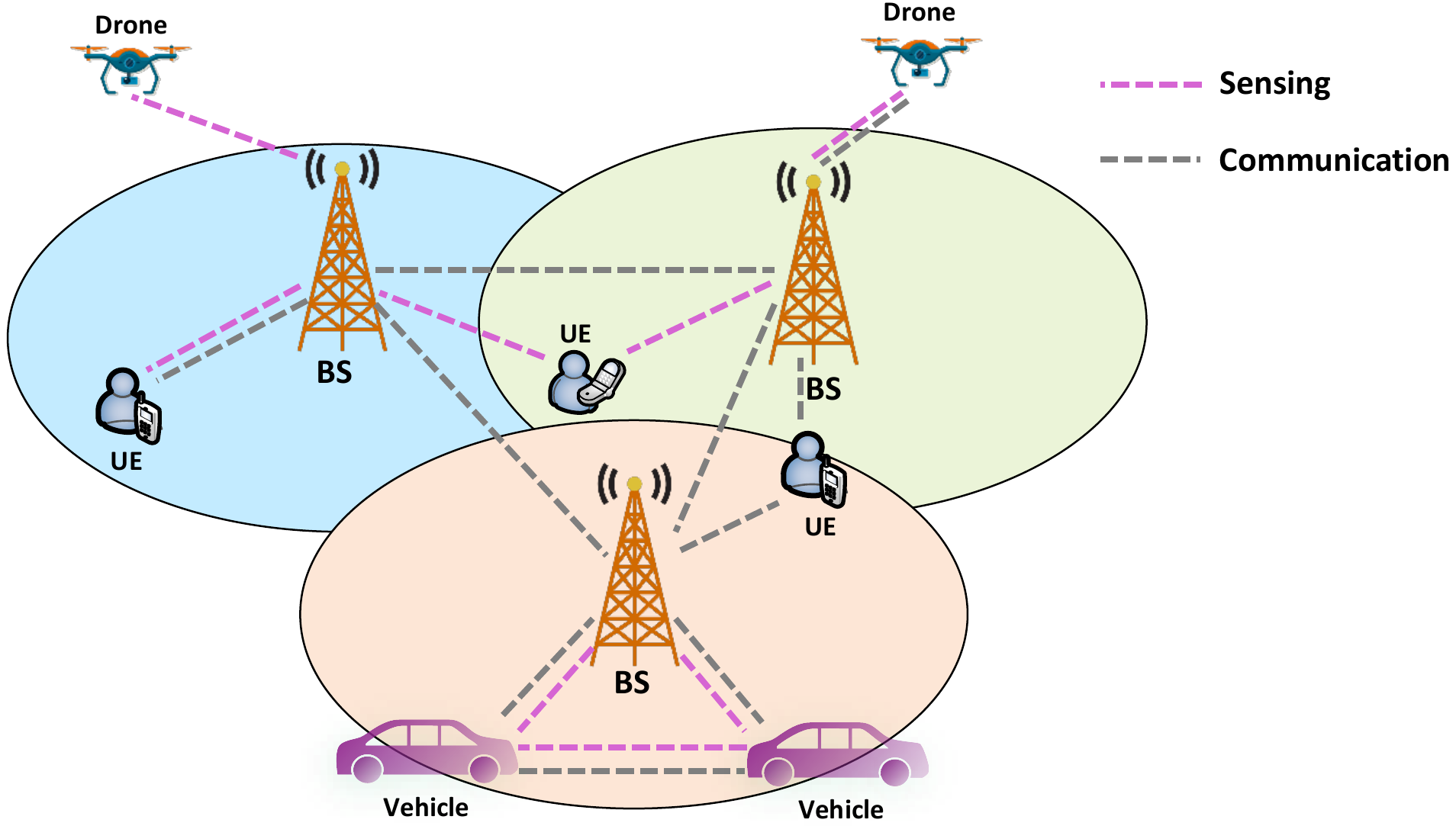}
            \caption{Illustration of a cellular network supporting native ISAC capabilities.}   \label{fig_ISAC_architecture}
\end{figure*}

Integrated sensing and communication (ISAC) has been identified as a promising technology for the next generation wireless communication, as well as a critical usage scenario of future communication systems \cite{liu_2023_6G}. ISAC will not only boost the experience for private users, but also support relevant vertical industries as well \cite{liu2023__6g}. Being able to provide sensing capabilities utilizing existing communication infrastructure, ISAC has advantages of lower costs, higher spectral efficiency and higher energy efficiency compared to counterparts that require dedicated spectrum and transceivers such as radar systems. A cellular network supporting ISAC as a native usage scenario is illustrated in Fig. \ref{fig_ISAC_architecture}, where sensing and communication functions are implemented in all base stations (BSs). These dual-functional BSs are able to support communication functions just like conventional BSs, while also able to utilize communication signals for sensing purposes. On top of getting data services, private users carrying user equipments (UEs), vehicles, drones and theoretically, any moving objects in the cell can be detected, tracked and localized by the BSs. It's envisaged that results obtained from sensing can help to improve communication features such as beamforming, and communication functions can also enhance the sensing accuracy. It is also possible for multiple BSs to jointly sense a common target, improving the resolution and accuracy of localization. Thanks to the propagation properties of wireless signals in free space, researchers have realized the potential of communication infrastructure in sensing and studied the theoretical and practical aspects of it \cite{9606831}. Even before the concept of ISAC, or sometimes called joint sensing and communication (JSAC), matures and becomes trendy,  there were work on utilizing communication signals for sensing and localization purposes \cite{5972832,8891248,9417324,8421080} which usually relied on classical signal processing techniques. 

As a critical step towards more in-depth research for ISAC systems, how to develop proper channel models draws increasing research attention. In \cite{9952200}, several channel modeling approaches for ISAC are summarized and a general procedure is recommended based on \cite{38901}. Specifically, the authors propose to introduce sensing clusters during channel modeling to better characterize the sensing process.
The authors of \cite{10078840} propose a channel modeling approach towards the 6th generation (6G) for ISAC, verifying their models with simulations results by comparing the analytical results, simulation results and measurement data.
Paving the way to mature the technology, studies have been conducted on different aspects such as electromagnetic modelling, optimal waveform design, joint beamforming \cite{9916163}, novel signal processing techniques \cite{9540344}, flexible duplex and control \cite{9967989} and achieving balance between accurate sensing and robust communication. 

These research findings build up the foundation of further prototyping and experiments. Coming to the practical side of the research, there are some literature on tests and trials of ISAC in different scenarios. \cite{7179321} introduces a gesture recognition system that leverages changes in WiFi signal strengths to sense in-air hand gestures around mobile devices. Communication links can be established simultaneously when conducting such trials.
The authors of \cite{9903001} develop an ISAC system using frame structure in the 4th generation (4G) communication system and orthogonal frequency division multiplexing (OFDM) signals. The experiments are conducted in an over-the-air (OTA) manner and reveal a graceful trade-off between the performance for communication and sensing.
In \cite{9162963}, an ISAC system for autonomous driving is designed and tested. The signals used for ISAC are designed according to the 5th generation (5G) communication standards, of 64 quadrature amplitude modulation (QAM) and cyclic prefix - OFDM (CP-OFDM). Signals are transmitted and received by tailored radio frequency (RF) feeds and receivers working on 28 GHz bands.
A recent work utilizes Wi-Fi signals for respiratory monitoring \cite{9989347}. The advantages of the system  include using commercial off-the-shelf (COTS) devices which can provide low-cost and simple respiratory monitoring without requiring specialized hardware. 
In \cite{ISAC_indoor_breath}, authors developed an indoor breath monitoring system using 5G BSs as the transmitters and receivers, and achieved satisfactory results. The tests were conducted under both line-of-sight (LOS) and non line-of-sight (NLOS) propagation conditions and the results were compared and analyzed. Recently, authors of \cite{10049817} developed an ISAC system that is tested in indoor lab environment with an automated driving vehicle. There are two ISAC systems deployed, distancing 2.7 meters from each other, and the vehicle is moving at a speed of 1 m/s. After data fusion, the positioning root mean square error (RMSE) is reported to be 0.1319 meters.

The idea of ISAC is also welcomed by researchers working on other promising technologies for the 6G networks, such as the reconfigurable intelligent surface (RIS) \cite{9679804,9955484}.
Authors of \cite{9133157} developed a RIS based communication system which is also capable of sensing human postures indoors. The system can distinguish four pre-defined human gestures using an optimized configuration. The system comprises of a pair of single antenna transceivers and uses tailored signal structures instead of standardized air interfaces.


To the best of the knowledge of the authors, there exists little work on deploying ISAC based approaches in 5G networks in practical outdoor environments. Thanks to the availability of 5G networks worldwide, even in remote areas \cite{8333695,9198057}, ISAC systems utilizing 5G have larger potentials in robustness than the ones relying on other signal sources. The wide range of available spectrum of 5G is also a huge advantage for sensing. What's more, since 6G is going to evolve from 5G and 5G-advanced, ISAC systems which are compatible to 5G can be readily upgraded to fit into future generation wireless networks upon the release of more design details of 6G. It is also rare to see experimental results of ISAC systems that are able to support detection, tracking and localization of multiple targets simultaneously.

Compared to related work, the prototype developed in this paper is unique in the sense of using standard 5G BSs with no hardware modifications, and the signals in the tests are all 5G signals capable of serving communication needs at the same time. It is also capable of conducting sensing for multiple targets simultaneously.

The rest of this paper is organized as follows. We describe the system model in Section \ref{Sec_model} and the approach to detect, track and localize targets in Section \ref{Sec_methods}.  In Section \ref{Sec_field_trial}, experimental results obtained in field trials are provided to demonstrate the feasibility and accuracy of the proposed method in practical outdoor environment. Finally, Section \ref{Sec:conclusion} concludes the paper.

\section{System Model}\label{Sec_model}
In this paper, a practical outdoor environment is considered. The goal is to utilize 5G BSs without any hardware modification to support sensing and communication simultaneously. The sensing capability is demonstrated by using 5G BSs to detect, localize and track interested targets in the test area. These targets are not required to be connected to the 5G network whatsoever.

The designed ISAC system works under bi-static mode, since there are two BSs involved in the process. One BS functions as the transmitter (Tx) and the other as the receiver (Rx). The two BSs are not required to be time synchronized. 
The Tx transmits 5G signals, whose frame structures and physical layer parameters are defined according to 5G standards \cite{38211} specified by the 3rd generation partnership project (3GPP). Note that to reap the integration gain of ISAC as much as possible, there is no modification to the frame structure so the transmitted signals can still be used for communication purposes.
Then, the reflected signals are received by the Rx. Note that since all the tests are conducted in fully practical environment, the propagation condition among the Tx, Rx and the targets are mixed line-of-sight (LOS) and non-LOS (NLOS). This will present some challenges to the experiments.

\begin{figure}[t] 
            \centering
            \includegraphics[width=8.8cm]{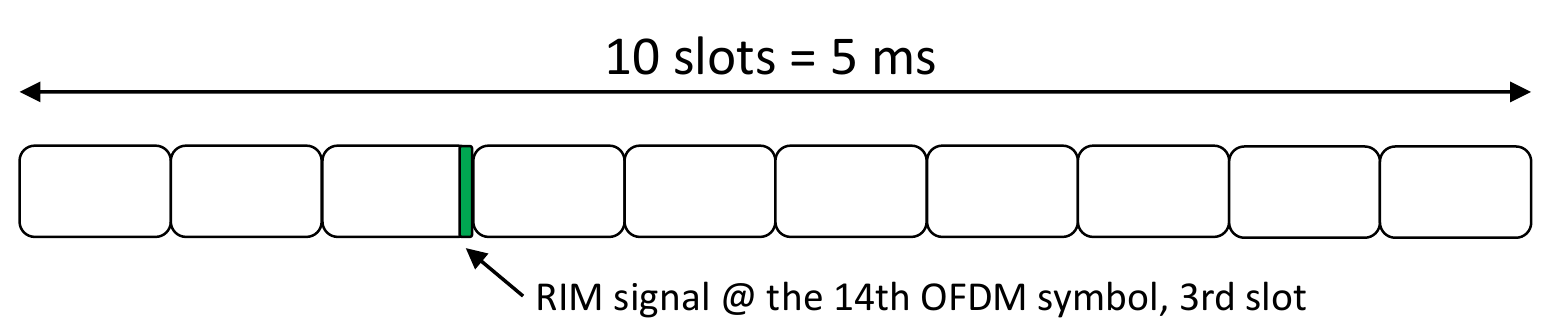}
            \caption{Frame structure of the test signal, where a half-frame is shown.}  \label{fig_frame}
\end{figure}

The frame structure of signals used in this paper is presented in Fig. \ref{fig_frame} where the subcarrier spacing is $30$ kHz and the length of each slot is $0.5$ ms. There are $10$ slots in a half-frame, which lasts $5$ ms. The signal used for target sensing is the remote interference management (RIM) signal, which is specified in \cite{38211} and placed on the 13th OFDM symbol of the 2rd slot. The periodicity of the RIM signal, which is used for sensing purposes, is $5$ ms, since one RIM signal is placed in a half-frame.

To make the system more general and applicable to most cases, there is few assumption on the targets. The targets can be small as a single pedestrian or big as a vehicle, while all need to be mobile. The constraint on the minimal mobility is introduced since there are many static objects surrounding the ISAC system such as buildings and trees, and it is necessary to filter these static objects out to enable detection and tracking of mobile targets.

\section{The ISAC Approach for Target Detection, Tracking and Localization}\label{Sec_methods}
Based on the system model introduced above, an ISAC system based on practical 5G networks is designed in this paper. The approach comprises of several steps, which are elaborated in details below. All the steps introduced below take place in the baseband, since it is assumed that decoding and demodulation are readily performed according to standard procedures specified by 3GPP.

\subsection{Data Pre-processing}
During the data pre-processing stage, it is necessary to perform delay and phase compensation as well as differential processing to obtain the time-domain signal segment for parameter estimation. Pre-processing can effectively improve the quality of the data and estimation accuracy in later stages. 
Assuming the input data consists of $T$ packets indexed by $t=0,…,T-1$. Each packet consists of frequency-domain received signals and has a size of $M \times N$, where $M$ represents the total number of sub-carriers for all resource blocks and $N$ denotes the number of antennas.
When the bandwidth for communication and sensing is 100 MHz, per the 3GPP standards, there are 273 resource blocks (RBs) and each RB has 12 carrier frequencies, which gives $M = 273 \times 12 = 3276$. 
In the trials, as the BSs are not synchronized and the target can be mobile, there are delay and phase deviations among different samples. To solve this problem, delay and phase compensation as well as differential processing need to be performed.

Firstly, it is necessary to process and store the data of the first packet. The data of the first packet contains the information on the multi-path generated from the environment as well as the ones generated by the targets. The phase of the other data packets can be aligned with the first data packet, and the multi-path components generated by the surrounding environment can be eliminated after differential processing. As a second step, it is necessary to calibrate the delay and phase deviations between different data packets by aligning the delay and phase. The difference among different packets is due to the potential mobility of the target being tracked. The phase calibration can be achieved by using the phase information extracted from the first packet to perform phase alignment for each data packet. At the same time, differential processing is used to calculate and calibrate the delay between adjacent data packets. The details are discussed below.

When receiving data packet indexed by $t ~ (t \neq 0)$, the correlation of this packet with the first packet is calculated as
\begin{align}
\mathbf{R}(0\!:\!M-1,N,t)\!=\!\mathbf{Y}(0\!:\!M-1,N,0)\mathbf{Y}(0\!:\!M-1,N,t)^H,
\end{align}
and the shift conjugate multiplication on the correlation matrix is calculated as
\begin{align}
\!\!\mathbf{A}(0\!:\!M\!-\!1,\!N\!,t)\!=\!\mathbf{R}(0\!:\!M-\!1-\!m_0,\!N\!,0)\mathbf{R}(m_0\!:\!M\!-\!1,\!N\!,t)^H,
\end{align}
where $m_0$ represents a default shift parameter.

The delay difference is calculated according to the phase difference as
\begin{align}
    \tau(N,t)=-\frac{\textup{angle}(\sum_{u=0}^{m-1}\mathbf{A}(u,N,t))}{m_0}.
\end{align}

To calculate the initial phase difference, it is required to perform delay difference compensation on the correlation matrix as
\begin{align}
    \mathbf{B}(0\!:\!M\!-\!1,\!N\!,t)=\mathbf{R}(0\!:\!M\!-\!1,\!N\!,0)\times\mathbf{D}^H,
\end{align}
where $\mathbf{D}=\exp{(j (-\frac{M}{2}:\frac{M}{2}-1)\otimes\tau(N,t))}$ with $\otimes$ representing the Kronecker product and the initial phase can be calculated as
\begin{align}
    P_{ini}(N,t) = \textup{angle}(\sum_{u=1}^{M-1}\mathbf{B}(u,\!N\!,t)).
\end{align}

The data after delay and phase compensation is
\begin{align}
    \mathbf{C}(0\!:\!M\!-\!1,\!N\!,t)=\mathbf{Y}(0\!:\!M\!-\!1,\!N\!,0)\times\mathbf{D}^H\times \mathbf{P}^H,
\end{align}
with $\mathbf{P}=\exp{(j\mathbf{1}_{M\times1}\otimes P_{ini}(N,t))}$.

Based on the above analysis, differential processing is performed on the delay-phase aligned data as
\begin{align}
    \mathbf{C}_{diff}(0\!:\!M\!-\!1,\!N\!,t)=\mathbf{C}(0\!:\!M\!-\!1,\!N\!,t)-\mathbf{C}(0\!:\!M\!-\!1,\!N\!,0),
\end{align}

To obtain the time-domain differential channel, the Ns-point inverse fast Fourier transform (IFFT) is conducted on the delay-phase aligned data as
\begin{align}
    \mathbf{h}_{diff}(0:N_s-1,N,t)=\textup{IFFT}(\mathbf{C}_{diff}(0\!:\!M\!-\!1,\!N\!,t),N_s).
\end{align}

Note that the channel response in the time domain is intercepted according to the desired detection range, and $T_s$ points are used for later processing steps.

\subsection{Doppler Estimation}
The Doppler estimation can be used to calculate the frequency shift caused by the Doppler effect during signal transmission and reception. Denote the steering vector as $\mathbf{S}(N,AoA_{id})$, where the angle of arrival (AOA) are defined as in \cite{wu2017favorable}. Here, $AoA_{id}$ represents the steering vector index whose dimension is $AoA_{H}\times AoA_{V}$. Moreover, $AoA_{H}$ and $AoA_{V}$ are the horizontal and azimuth angles of the detection, respectively, and the values are determined by the desired detection range and step size.

To carry out the Doppler estimation, firstly, within the time range corresponding to the total $T$ packets, the processing results $\mathbf{h}_{diff}(T_s,N,t)$ can be generated from a given packet $t$ according to the order of signal arrival. Since all packets need to be processed consistently, it is necessary to save the output of all packets and record the saved result as $\mathbf{h}_{diff}(T_s,N,1:T)$.

In the next step, $T$-point fast Fourier transform (FFT) is performed on the collected data to obtain the Doppler shift as
\begin{align}
    \textup{Doppler}(T_s,N,D_{id})=\textup{FFT}(\mathbf{h}_{diff}(0:T_s,N,1:T)),
\end{align}
where $D_{id}$ represents the Doppler frequency domain index.

For each point in the delay-Doppler domain, AOA spectral peak search is performed to obtain the maximum spectral peak and the corresponding horizontal and azimuth angle index. First, the data in the delay-Doppler domain with the steering vector is correlated to obtain the correlation power as
\begin{align}
    \textup{Cor}(t_S,AoA_{id},D_{id})\!\!=\!\!|\textup{Doppler}(t_S,N,d_{id})\times \mathbf{S}(N,AoA_{id})|^2 .
\end{align}
Then, for all delay-Doppler-domain points, the maximum of the correlation power value is searched. That particular value and the corresponding index are recorded and $v_{peak}, AoA_{id}, d_{id}$ can be obtained. For the part near zero Doppler, which represents objects with very mobility, the AOA peak search can be omitted to reduce computational complexity since the interested targets are assumed to have a certain level of mobility.

\subsection{Target Selection}
In the delay-Doppler domain, the target selection is used to search for target peaks. To achieve that, the following steps are performed in the algorithm.

In step 1, delay partitioning is executed. In the delay dimension, each region comprises of $m$ consecutive $T_s$ points. There is a total of $N_c$ regions, and all peaks are searched.

In step 2, the peaks obtained from step 1 are sorted in the descending order by their power and the top $k$ targets are selected. The same operation is performed on all delay partitions to obtain $kN_c$ targets.

In step 3, a total number of $kN_c$ individual targets are sorted by power in descending order and the top $U$ targets with energy greater than the pre-defined threshold are selected. These $U$ targets are the ones identified at this particular snapshot and will be given to the next step for tracking.

\subsection{Target Tracking}
At time $t$, the number of occurrences for each new target is recorded as $1$. At time $t+1$, for each existing target, the target which has minimum Euclidean distance that less than the threshold is found, and the number of occurrences corresponding to this target is increased by one. Targets that appear in $t+1$ but not in $t$ are considered new targets.

However, to decrease the false alarm rate, new targets are not directly considered as formal targets at the first appearance. A new target is considered a formal target only if it appears for $p$ consecutive times with $p$ being the threshold. At the same time, for each target appearing in $p$ consecutive samples, the average velocity of the previous $p-1$ points must be larger than the minimum velocity $v_{min}$. Otherwise, the target will not be identified and tracked. Since only moving objects are interested, setting a threshold on the velocity can help to reduce computational complexity by reducing the number of targets need to be tracked.

\subsection{Target Localization}
As a next step after identifying and tracking the targets, determining the coordinates of the targets comes into play. 
For the targets, their coordinates are calculated respectively based on the corresponding appearance time, horizontal angle and azimuth angle information.

The receiving BS is assumed to be located in the origin point. For a particular target, denote $(x, y, z)$ as its coordinates, $(a, b, c)$ as the coordinates of the transmitting BS, $r'$ as the distance between the target and the receiving antenna, $\theta$ as the elevation angle, and $\psi$ as the azimuth angle. Moreover, the sum of the distance between the transmitting BS to the target and the distance between the target and the receiving BS, which is denoted $r$, is given by
\begin{align}
    r = r' + \sqrt{(x-a)^2+(y-b)^2+(z-c)^2},
\end{align}
where
\begin{align}
    x&=r'\sin\theta\cos\psi,\\
    y&=r'\sin\theta\sin\psi,\\
    z&=r'\cos\theta.
\end{align}


Thus, $r'$ can be re-written as
\begin{align}
    r' = \frac{a^2+b^2+c^2-r^2}{2a\sin\theta\cos\psi+2b\sin\theta\sin\psi+2c\cos\theta-2r},
\end{align}
and the Cartesian coordinate of the target can be obtained. Combined with target tracking strategy, the location information of mobile target users can be renewed.

\section{Experimental Results}\label{Sec_field_trial}
In this Section, the results obtained from a group of field trials are presented to demonstrate the feasibility and accuracy of the proposed ISAC based multi-target detection and tracking approach. 

The BSs used in all tests are 5G BSs without any hardware modification and are configured in a normal working mode. UEs passing through the test area can normally connect to 5G network during all tests.

There are in total of four trials conducted in this paper, which will be elaborated below in sequential order.

While striving to test the performance of communication on top of sensing, it is very challenging to categorize and test communication performance in practical networks since the BSs used in the trials are up-and-running 5G BSs and it's complicated to evaluate the communication performance since it depends on the users nearby. Theoretically, communication functions shouldn't be interfered since the sensing signal is embedded into the frame structures of 5G signals according to 3GPP standards.

\subsection{General Test Setup}
\begin{figure*}
            \centering
            \includegraphics[width=17cm]{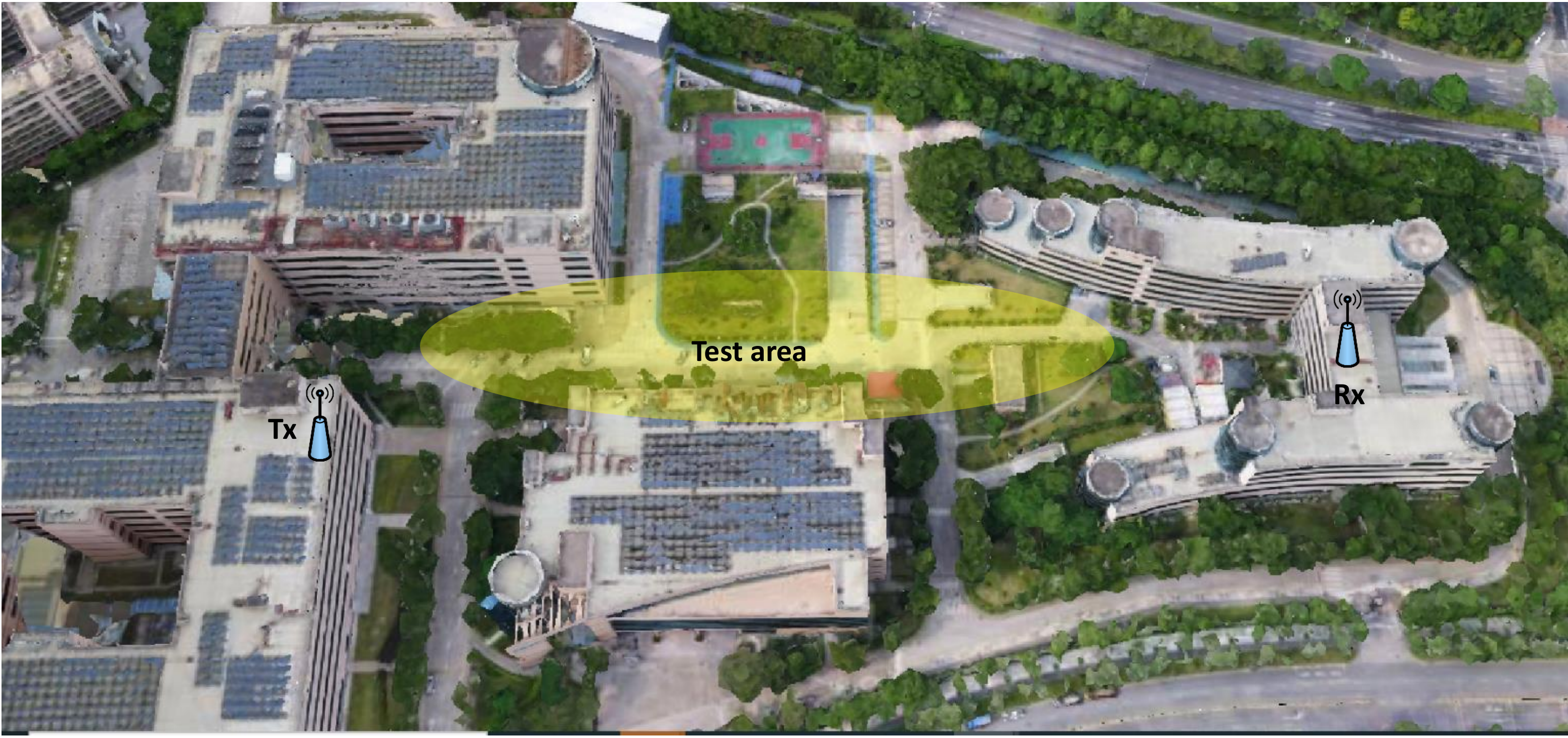}
            \caption{The test area.}   \label{fig_main_scenario}
\end{figure*}

To fully test the proposed approach and the system, comprehensive trials are conducted in a practical environment located in the ZTE Xili industrial park, Shenzhen, China. As depicted in Fig.~\ref{fig_main_scenario}, there are 2 BSs involved in the tests, serving as the Tx and Rx, respectively. Both BSs are on the top of the buildings, overlooking the test area in between. The detailed configuration of the BSs are given in Table \ref{Table_config}, which follows standard 3GPP specifications.
As the trials are conducted in practical environment,  there are buildings, trees, grassland, parked vehicles and pedestrians in the test area, imposing shadowing and fading effects. 
During all trials, there is no restriction on the traffic in the test area, which means there can be moving objects such as vehicles and pedestrians. The test results are demonstrated by detecting and tracking intended targets, such as drones controlled by the authors, while other objects nearby are also visible to the ISAC system and can also be detected and tracked.


\begin{table}[t]
  \caption{Configurations of the transmitting and receiving BS}
  \label{Table_config}
  \centering
  \begin{tabular*}{0.55\columnwidth}{l p{0.15\columnwidth}}
  \hline\hline
  Parameter            & Value \\
  \hline
  center frequency   &  $4850$ MHz \\
  bandwidth         &  $100$ MHz \\
  subcarrier spacing   &  $30$ kHz \\
  number of Tx antennas         &  64 \\
  number of Rx antennas         &  64 \\
  waveform         &  OFDM \\
  \hline\hline
  \end{tabular*}
\end{table}

\subsection{Detection and Localization of A Stepping Pedestrian}
As a first step, a simple trial is carried out to detect and localize a pedestrian stepping at a specific location. The ISAC system detects the pedestrian and estimates its position. To evaluate the estimation accuracy, the pedestrian carries a global positioning system (GPS) tracker to obtain a benchmark position. The tests are conducted in 3 different locations in the test area and the results are given in Table \ref{Table_stepping_man}.

\begin{table*}
  \caption{Localization results of a stepping pedestrian}
  \label{Table_stepping_man}
\centering
\begin{tabular}{|p{18pt}|p{110pt}|p{110pt}|p{33pt}|}
\hline
Test    &  Estimated coordinates   & GPS coordinates & Error (m) \\
\hline
  1    &  (113.9302430 E, 22.58280501 N)     & (113.9302483 E, 22.58280606 N) & 0.55 \\
\hline
  2   &  (113.9300548 E, 22.58280501 N)    & (113.9300669 E, 22.58280335 N) &  1.26 \\
  \hline
  3    &  (113.9301468 E, 22.58280501 N)   & (113.9301580 E, 22.58280619 N) & 1.15 \\
\hline
\end{tabular}
\end{table*}

The distance from the three locations to the BSs are around $100$ to $200$ meters. As can be seen in Table \ref{Table_stepping_man}, the average error of localizing the stepping pedestrian is  $0.99$ m. Considering the large area where the pedestrian can appear as well as the practical outdoor environment where other moving objects exist, such an accuracy can be satisfactory.

\subsection{Tracking of A Walking Pedestrian}

\begin{figure}
            \centering
            \includegraphics[width=0.98\linewidth]{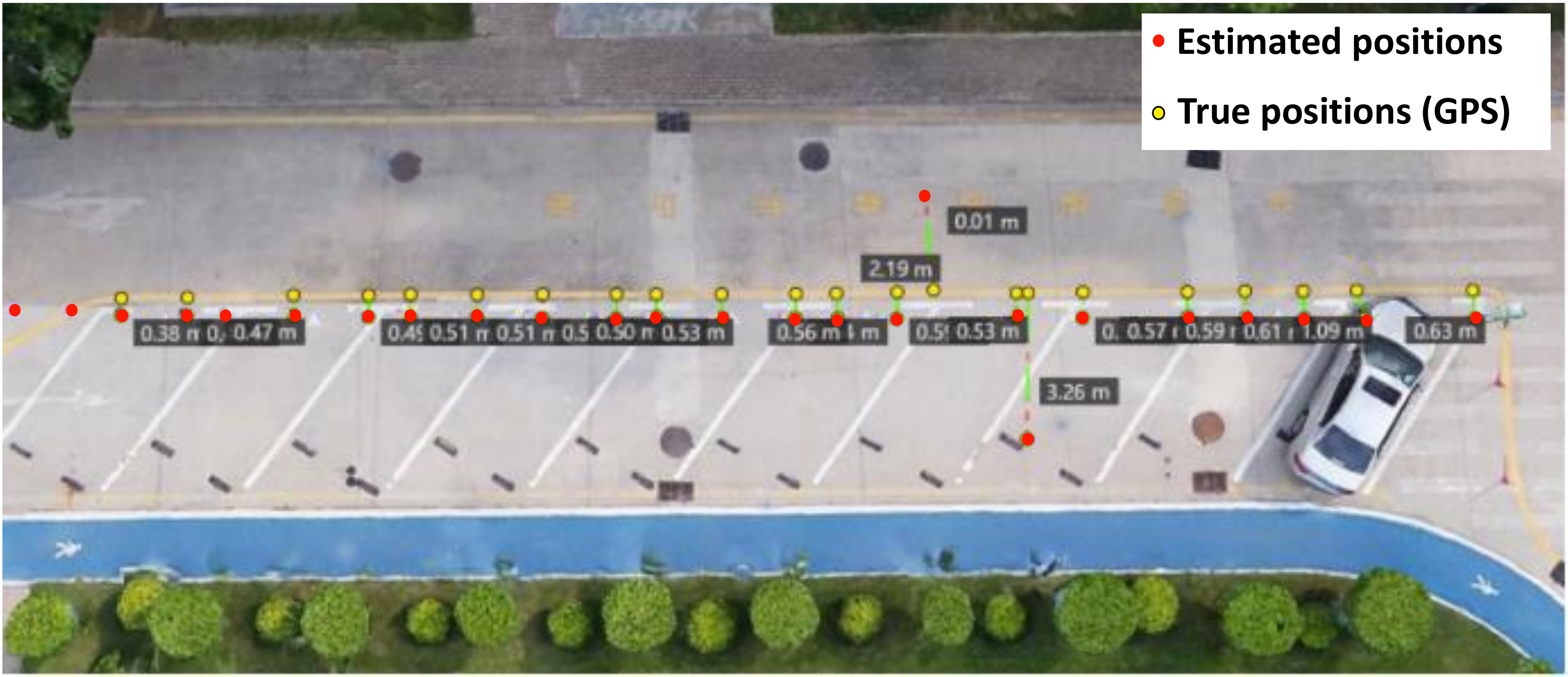}
            \caption{Tracking of a walking pedestrian, where the GPS coordinates are depicted in yellow dots, estimated positions are represented by red dots and errors are marked in tags.} \label{fig_walking_man}
\end{figure}

To test the performance of tracking moving objects, a walking pedestrian is tracked using the ISAC system. As a human being is relatively a very small object and presents only small changes in channel responses, tracking a walking pedestrian is considered challenging, compared to tracking larger objects such as vehicles. As shown in Fig.~\ref{fig_walking_man}, the test scenario is set at a small parking lot alongside the street, where one car is parked and remains static. The pedestrian walks at a regular pace across the test field, from one side to the other.  To demonstrate the real accuracy of the ISAC system, the estimations on positions are based on data acquired from every single shot, without considering history positions or applying any filtering approaches to smooth the estimated trajectory.

\begin{figure}
            \centering
            \includegraphics[width=0.98\linewidth]{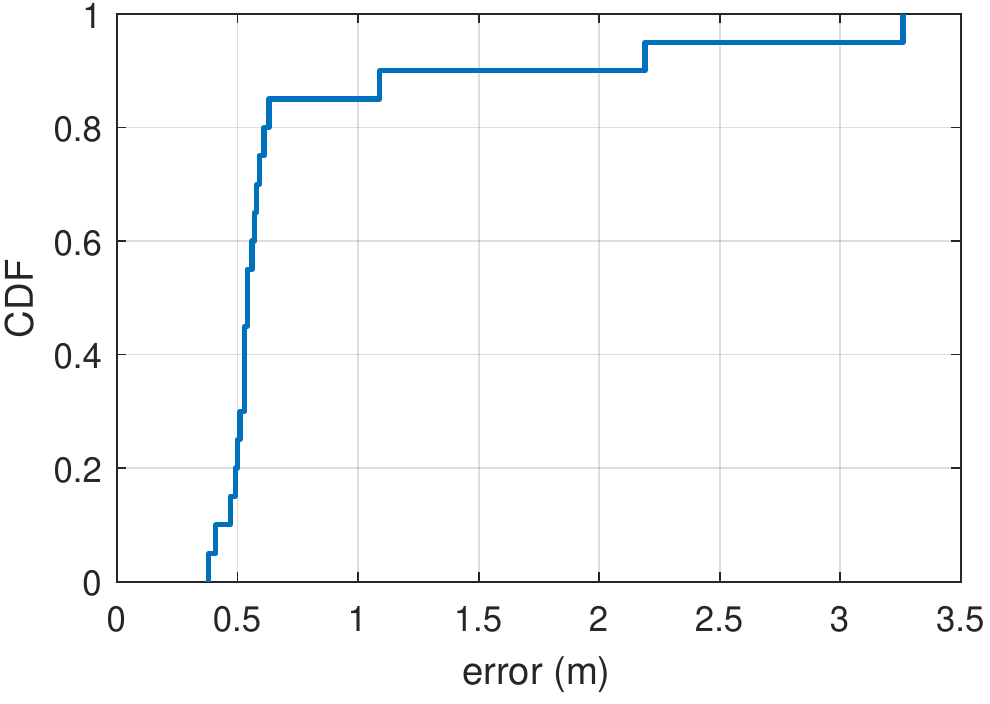}
            \caption{Cumulative distribution function of the errors in tracking the location of a walking pedestrian.} \label{fig_cdf}
\end{figure}

\begin{figure*}
            \centering
            \includegraphics[width=15cm]{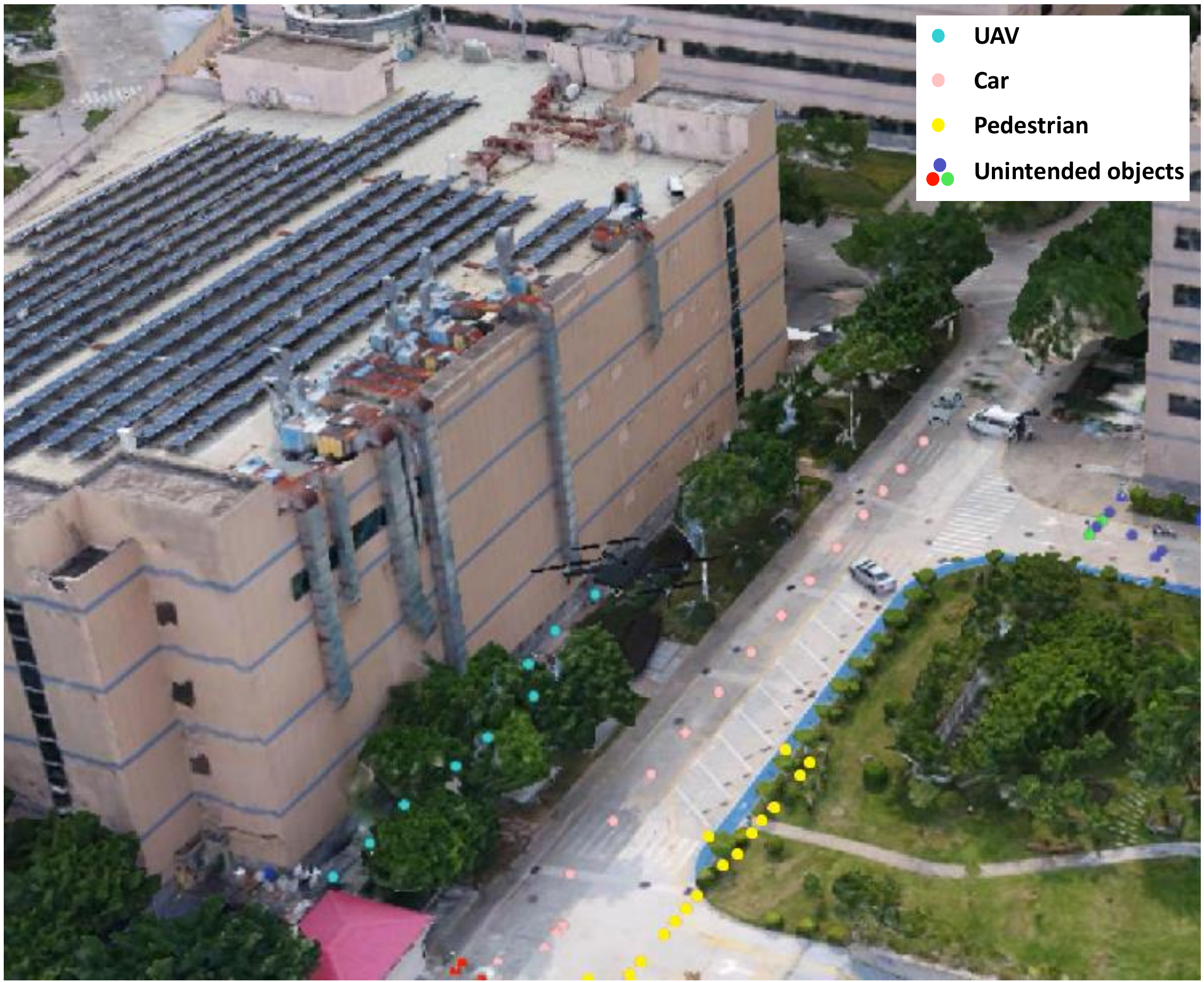}
            \caption{Test environment and results of simultaneous multi-target detection and tracking.}   \label{fig:multi_target}
\end{figure*}

The errors of estimation, calculated by the distance from the GPS coordinates to the corresponding estimation at the same time slot, are given by values in tags in Fig.~\ref{fig_walking_man}. During the test comprised of 20 snapshots, the majority of estimations give a quite satisfactory accuracy with estimation error less than 0.6 m, as depicted in Fig.~\ref{fig_cdf}. Notably, there are two estimations with significantly larger errors, specifically, 2.19 m and 3.26 m. This is potentially due to random interference presented in the radio environment at that time. The average localization error is 1.03 m and 0.58 m throughout the test, respectively, taking into account of the two outliers or not. 
It is also noted that within all estimations, the minimum error is 0.38 m. The accuracy of tracking can be further improved by taking history data into consideration, such as applying the Kalman filter, to smooth the trajectory and minimize the impact of potential outliers.

\subsection{Simultaneous Tracking of UAV, Car and Pedestrian}
Theoretically, the ISAC system built on 5G networks can detect and track multiple targets simultaneously, while this capability is rarely demonstrated in previous work. In this section, multi-target detection and tracking is performed in the test area, with three objects of different reflective characteristics, altitude and mobility, a pedestrian, a car and an unmanned aerial vehicle (UAV). The test environment as well as the results are shown in Fig. \ref{fig:multi_target}. The UAV cruises from one end of the test area to the other at a speed of approximately 3 m/s while maintaining the height of 30 meters, as depicted by the light blue dots. The car crosses the test area also at a constant speed of 3 m/s, as represented by the pink dots. The pedestrian walks alongside the edge of the green who can be identified by yellow dots. As the test is conducted in an outdoor open area, there can always be other people walking around who enters the test area accidentally. The unwanted appearance is also detected by the ISAC system and tracked as well, as depicted by the dark blue, red and green dots. 

The positioning errors of the three intended objects all present a similar distribution to the results when tracking a single target, as  depicted in Fig. \ref{fig_cdf}. The minimum error in the three dimensional space is roughly 0.3 m and the largest error is around 3 meters. 
The results demonstrate that the designed ISAC system in this paper can track and localize multiple target simultaneously without loss of the localization accuracy compared to the case of tracking a single target.
How to eliminate outliers is still one of the key to decrease the average positioning error and improve the accuracy. 

\subsection{Long Range UAV Tracking}

\begin{figure*}[t]
            \centering
            \includegraphics[width=16cm]{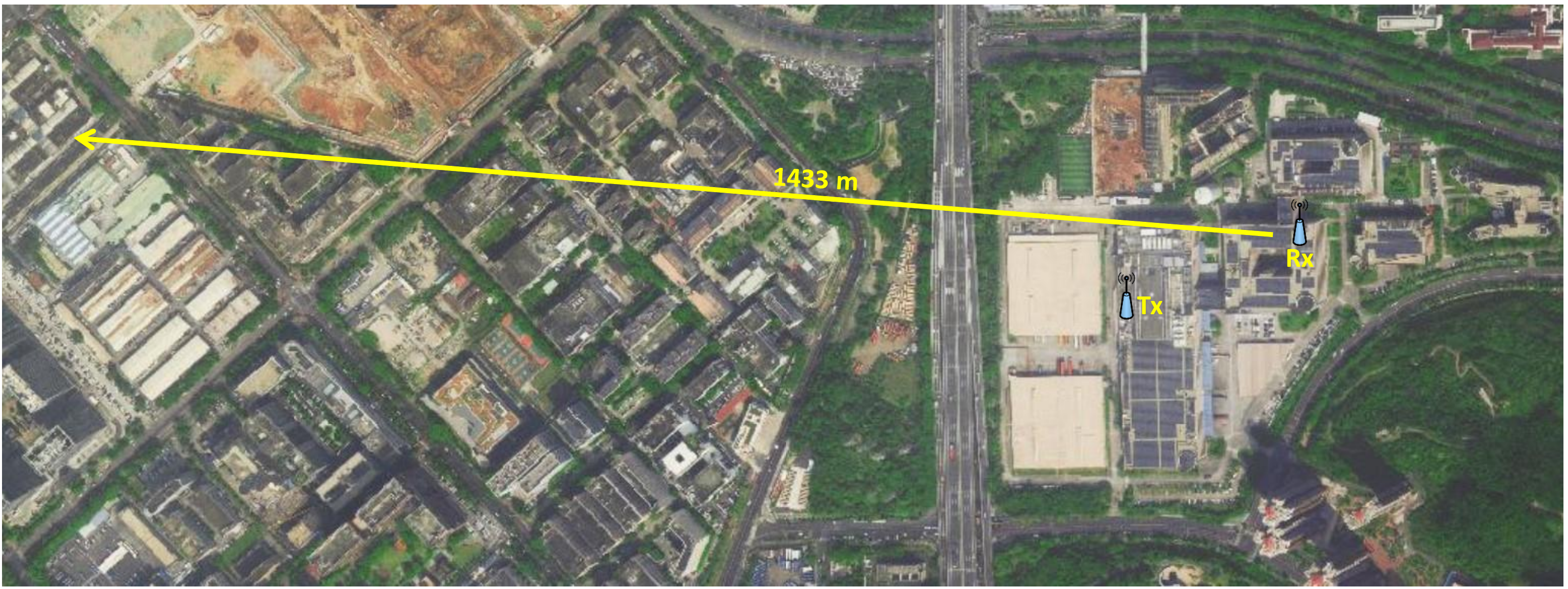}
            \caption{The long range UAV tracking test.}   \label{fig_large_map}
\end{figure*}

On top of testing the accuracy of the proposed ISAC system, the effective area of it is also of great interest and thus needs to be verified. The effective area can be complicated to describe, as it depends on the geometrical settings of the surrounding environment. To simplify, a largest detection and tracking distance is proposed to be used as a metric to represent the capability of the ISAC system in terms of detecting and tracking targets far away. Note that the transmitting BS used in the test does not have any extra power boosting. 

To test the largest detection and tracking distance of the proposed ISAC system, another trial is conducted in the same industrial park, as shown in Fig.~\ref{fig_large_map}. The cruising path of the drone is set according to local regulations for flying UAVs, taking off from the industrial park and heading west. To better track the drone on this direction, two different BSs are used as Tx and Rx in this particular test, which are still 5G BSs without any hardware modifications. The coordinates of the Tx is (113.927001 E, 22.581170 N, 61.20 m) and the Rx is located at (113.929047 E, 22.582672 N, 64.39 m), as measured by the GPS.
The two BSs still serve UEs nearby as normal network access nodes during the whole test.

The largest distance is measured from the starting point where the UAV takes off near the Rx, to the ending point where the ISAC system cannot track and detect the target anymore. The maximal detection distance achieved in this test is 1433 meters. Considering that the 5G network is deployed in a relatively dense way of approximately 100 BSs in 1 square kilometer (depending on the environment and scenarios), such a tracking distance can empower 5G networks the capability to serve as a sensing network across the whole deployment area.


\section{Conclusion}\label{Sec:conclusion}
In this paper, world's first ISAC system empowered by practical 5G network in outdoor environment is designed and tested. The proposed system uses 5G BSs serving nearby mobile users normally to enable high-accuracy ISAC, which can achieve the positioning accuracy of approximately 1 meter in a test area of hundreds of meters long and wide. Moreover, the system is capable of detecting, tracking and localizing multiple targets simultaneously, as demonstrated by tracking a pedestrian, a car and an UAV, without loss of localization accuracy compared to the case of tracking a single target.
It is also verified that by using 5G BSs without any extra power boosting, the ISAC system can achieve a tracking range of more than 1.4 km. The experimental results obtained in the field trials confirm the feasibility of supporting ISAC using 5G networks, and pave way for future research and engineering of practical ISAC systems.

\bibliographystyle{IEEEtran}
\bibliography{Reference}

\begin{IEEEbiography}[{\includegraphics[width=1in,height=1.25in,clip,keepaspectratio]{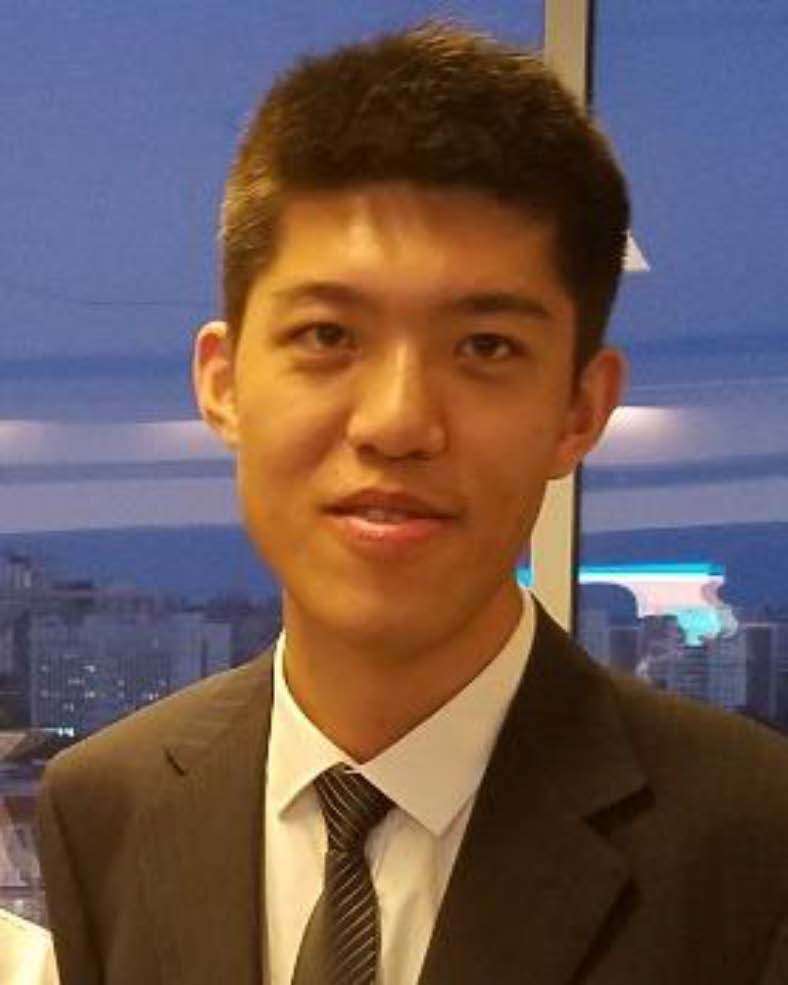}}]{Ruiqi (Richie) Liu }  (Member, IEEE) received the B.S. and M.S. degree (with honors) in electronic engineering from the Department of Electronic Engineering, Tsinghua University in 2016 and 2019 respectively. He is now a master researcher in the wireless research institute of ZTE Corporation, responsible for long-term research as well as standardization. His research interests include reconfigurable intelligent surfaces, integrated sensing and communication, wireless positioning and quantum communication. He is the author or co-author of several books and book chapters. During his 3-year service at 3GPP from 2019 to 2022, he has authored and submitted more than 500 technical documents with over 100 of them approved, and he served as the co-rapporteur of the work item (WI) on NR RRM enhancement and the feature lead of multiple features. He currently serves as the Vice Chair of ISG RIS in the ETSI. He actively participates in organizing committees, technical sessions, workshops, symposia and industry panels in IEEE conferences as the chair, organizer, moderator, panelist or invited speaker. He served as the guest editor for Digital Signal Processing and the lead guest editor for the special issue on 6G in IEEE OJCOMS. He serves as the Editor of ITU Journal of Future and Evolving Technologies (ITU J-FET) and the Associate Editor of IET Quantum Communication. He is the Standardization Officer for IEEE ComSoc ETI on reconfigurable intelligent surfaces (ETI-RIS) and the Standards Liaison Officer for IEEE ComSoc Signal Processing and Computing for Communications Technical Committee (SPCC-TC). He received the Outstanding Service Award from the SPCC-TC in 2022.
\end{IEEEbiography}

\begin{IEEEbiography}[{\includegraphics[width=1in,height=1.25in,clip,keepaspectratio]{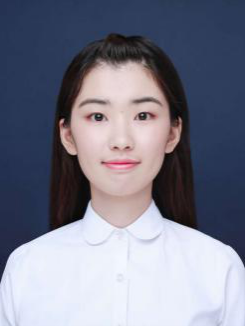}}]{Mengnan Jian } received the B.E. degree in information engineering from the Beijing Institute of Technology, Beijing, China, in 2016 and the M.S. degree from Tsinghua University, Beijing, China, in 2019. She is currently an engineer in ZTE corporation. Her research interests include reconfigurable intelligent surfaces, holographic MIMO and orbital angular momentum.
\end{IEEEbiography}

\begin{IEEEbiography}[{\includegraphics[width=1in,height=1.25in,clip,keepaspectratio]{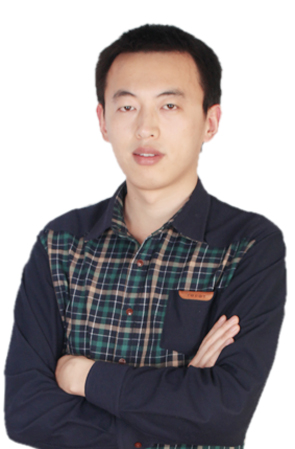}}]{Dawei Chen } received the B.S. degree from Northeast Agricultural University in 2012 and the the M.S. degree from Harbin Institute of Technology in 2015. Now he is a senior algorithm engineer of ZTE, with a research focus on integrated sensing and communication and high-precision indoor positioning. He has applied for more than 30 patents and published more than 20 papers.
\end{IEEEbiography}

\begin{IEEEbiography}[{\includegraphics[width=1in,height=1.25in,clip,keepaspectratio]{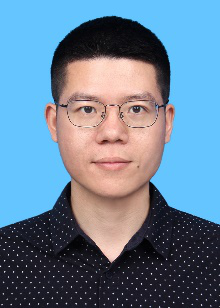}}]{Xu Lin } received the B.S. degree from  Harbin Institute of Technology (HIT), Harbin, China, in 2014, the M.S. degree from the China Academy of Telecommunications Technology, Beijing, China, in 2017, and the Ph.D. degree from  HIT, Harbin, China, in 2022. From 2019 to 2020, he was a Research Trainee with the Department of Electrical and Computer Engineering, McGill University, Montreal, Canada. He is currently an Algorithm Engineer with ZTE Corporation, and he is also a Postdoctoral Research Fellow under the joint supervision of ZTE Corporation and HIT. His research interests include communication signal processing, physical waveform design, transform domain communication systems, and integrated sensing and communication.
\end{IEEEbiography}

\begin{IEEEbiography}[{\includegraphics[width=1in,height=1.25in,clip,keepaspectratio]{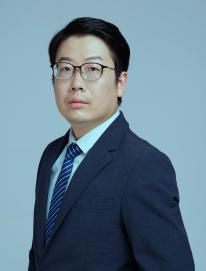}}]{Yichao Cheng } received the B.S. degree from University of Electronic Science and Technology of China in 2005. He is currently an engineer in ZTE Corporation. His research interests include 5G-Advanced software architecture design and 5G-Advanced integrated sensing and communication.
\end{IEEEbiography}

\begin{IEEEbiography}[{\includegraphics[width=1in,height=1.25in,clip,keepaspectratio]{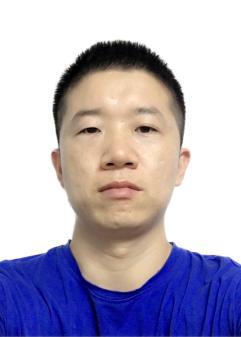}}]{Wei Cheng } received his Master's degree from Central South University, China in 2011. He is now an algorithm engineer with the wireless research institute, ZTE Corporation. His research interests include wireless communication and integrated sensing and communication.
\end{IEEEbiography}

\begin{IEEEbiography}[{\includegraphics[width=1in,height=1.25in,clip,keepaspectratio]{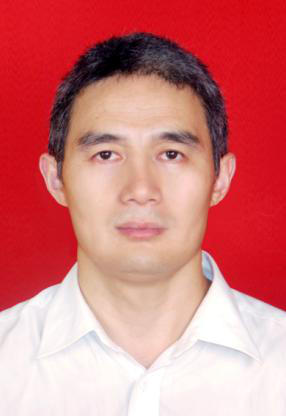}}]{Shijun Chen } graduated from Harbin Engineering University in 1999 with a master's degree. He is now an algorithm engineer with the wireless research institute, ZTE Corporation.  His current research interests include wireless positioning, wireless perception, channel simulation and industrial software. He has won 10 provincial and municipal science and technology awards. He has applied for more than 100 patents and published over 30 papers.
\end{IEEEbiography}

\end{document}